\documentclass[vecphys]{svmult}

\usepackage{makeidx}         
\usepackage{graphicx}        
\usepackage{multicol}        
\usepackage{amsmath}

\makeindex             

\begin{document}

\title*{
Fast polaron  switching in degenerate molecular quantum dots}
\titlerunning{Polaron switching in molecular devices}

\author{A. S. Alexandrov$^1$  and A. M. Bratkovsky$^2$ }
\institute{ $^1$ Physics Department, Loughborough University,
Loughborough LE11 3TU, United Kingdom\\ $^2$Hewlett-Packard
Laboratories, 1501 Page Mill Road, Palo Alto, California 94304,
United States }
%
%
\maketitle

Devices for nano- and molecular size electronics are currently a
focus of research aimed at an efficient current rectification and
switching.  Current switching due to conformational changes in the
molecules is slow, on the order of a few kHz. Fast switching ($\sim
$1~THz) may be achieved, at least in principle, in a degenerate
molecular quantum dot with strong coupling of electrons with
vibrational excitations. We show that the mean-field approach fails
to properly describe intrinsic molecular switching and present an
exact solution to the problem.

\section{Introduction}


For many applications one needs an {\em intrinsic} molecular
``switch'', i.e. a bistable voltage-addressable molecular system
with very different resistances in the two states that can be
accessed very quickly \cite{donhauser}. There is a trade-off between
the stability of a molecular state and the ability to switch the
molecule between two states with an external perturbation (we
discuss an electric field, switching involving absorbed photons is
impractical at a nanoscale). Indeed, the applied electric field, on
the order of a typical breakdown field $E_{b}\leq 10^{7}$V/cm, is
much smaller than a typical atomic field $\sim 10^{9}$V/cm,
characteristic of the energy barriers. Small barrier would be a
subject for sporadic thermal switching, whereas a larger barrier
$\sim 1-2$eV would be impossible to overcome with the applied field.
One may only change the relative energy of the minima by external
field and, therefore, redistribute the molecules statistically
slightly inequivalently between the two states. An intrinsic
disadvantage of the conformational mechanism \cite{ReedRAM},
involving motion of ionic group, exceeding the
electron mass by many orders of magnitude, is a slow switching speed ($\sim $%
kHz). In case of supramolecular complexes like rotaxanes and
catenanes \cite {rotax} there are two entangled parts which can
change mutual positions as a result of redox reactions (in
solution). Thus, for the rotaxane-based memory devices a slow
switching speed of $\sim 10^{-2}$ seconds was reported.

We have, therefore, explored a possibility for a fast molecular
switching where switching is due to strong correlation effects on
the molecule itself, so-called molecular quantum dot (MQD).
\begin{figure}[ptb]
\begin{center}
\includegraphics [angle=0, width=0.7\textwidth]
{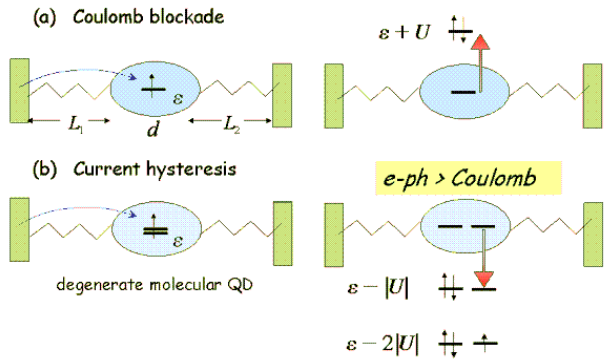}
\end{center}
\caption{Schematic of the molecular quantum dot with central
conjugated unit separated from the electrodes by wide-band
insulating molecular groups. First electron tunnels into the dot and
occupies an empty (degenerate) state there. If the interaction
between the first and second incoming electron is repulsive, $U>0$,
then the dot will be in a Coulomb blockade regime (a). If the
electrons on the dot effectively attract each other, $U<0$, the
system will show current hysteresis (b).} \label{fig:moldot}
\end{figure}
The molecular quantum dot consists of a central {\em conjugated}
unit (containing half-occupied, and, therefore, extended $\pi -$orbitals), Fig.~%
\ref{fig:moldot}. Frequently, those are formed from the p-states on
carbon atoms, which are not {\em saturated} (i.e. they do not share
electrons with other atoms forming strong $\sigma -$bonds, with
typical bonding-antibonding energy difference about 1Ry). Since the
$\pi -$orbitals are half-occupied, they form the HOMO-LUMO states.
The size of the HOMO-LUMO gap is then directly related to the size
of the conjugated region $d$, Fig.~\ref {fig:moldot}, by a standard
estimate $E_{\text{HOMO-LUMO}}\sim \hbar ^{2}/md^{2}\sim 2-5$~eV. It
is worth noting that in the conjugated linear polymers like
polyacetylene ($-\stackrel{|}{C}=\stackrel{|}{C}$)$_{n}$ the
spread of the $\pi -$electron would be $d=\infty $ and the expected $E_{%
\text{HOMO-LUMO}}=0.$ However, such a one-dimensional metal is
impossible, Peierls distortion (C=C bond length dimerization) sets
in and opens up a gap of about $\sim 1.5$eV at the Fermi level
\cite{yong03,DSindSw04,molswitchbook01}. In a molecular quantum dot
the central conjugated part is separated from electrodes by
insulating groups with saturated $\sigma -$bonds, like e.g. the
alkane chains, Fig.~3. Now, there are two main possibilities for
carrier transport through the MQD. If the length of at least one of
the insulating groups $L_{1(2)}$ is not very large (a conductance
$G_{1\text{(}2)}$ is not much smaller than the conductance quantum
$G_{0}=2e^{2}/h)$, then the transport through the MQD will proceed
by resonant tunneling processes. If, on the other hand, both groups
are such that the tunnel conductance $G_{1(2)}\ll G_{0},$ the charge
on the dot will be quantized. Then we will have another two
possibilities:
(i) the interaction of the extra carriers on the dot is {\em repulsive} $%
U>0, $ and we have a Coulomb blockade \cite{CoulBlock}, or (ii) the
effective interaction is {\em attractive}, $U<0,$ then we would
obtain the current {\em hysteresis} and switching \cite{AB03vibr}
(see below). Coulomb blockade in molecular quantum dots has been
demonstrated in Refs.~\cite{molSET}. In these works, and in
Ref.~\cite{zhitvibr}, the three-terminal active molecular devices
have been fabricated and successfully tested.

Much faster switching compared to the conformational one may be
caused by coupling to the vibrational degrees of freedom, if the
vibron-mediated attraction between two carriers on the molecule is
stronger than their direct Coulomb repulsion,
Fig.~\ref{fig:moldot}b. The attractive energy (i.e. a negative
"Hubbard" $U$)  is the difference of two large interactions, the
Coulomb repulsion and the phonon mediated attraction, on the order
of $1{\rm eV}$ each, hence $|U|\sim 0.1$eV.

\section{ Failure of mean field model of polaron molecular
switching }

Although the correlated electron transport through mesoscopic
systems with repulsive electron-electron interactions received
considerable attention in the past, and continues to be the focus of
current studies, much less has been known about a role of
electron-phonon correlations in ``molecular quantum dots'' (MQD).
Some while ago we have proposed a negative$-U$ Hubbard model of a
$d$-fold degenerate quantum dot \cite{alebrawil} and a polaron model
of resonant tunneling through a molecule with degenerate level
\cite{AB03vibr}. We found that the \emph{attractive} electron
correlations caused by any interaction within the molecule could
lead to a molecular \emph{switching} effect where I-V
characteristics have two branches with high and low current at the
same bias voltage. This prediction has been confirmed and extended
further in our theory of \textit{correlated} transport through
degenerate MQDs with a full account of both the Coulomb repulsion
and realistic electron-phonon (e-ph) interactions. We have shown
that while the phonon side-bands significantly modify the shape of
hysteretic I-V curves in comparison with the negative-$U$ Hubbard
model, switching remains robust. It shows up when the effective
interaction of polarons is attractive and the state of the dot is
multiply degenerate, $d>2$.

Nevertheless, later on Galperin \textit{et al}.~\cite{ratsw} argued,
without discussing the discrepancies with the prior work, that even
a non-degenerate electronic level ($d=1)$ coupled to a single
vibrational mode produces a hysteretic I-V curve, a current
switching, and a negative differential resistance. Here we
explicitly calculate I-V curves of the nondegenerate ($d=1)$ and
two-fold degenerate ($d=2)$ MQDs to show that these findings are
artefacts of the mean-field approximation used in Ref. \cite{ratsw}
that neglects the Fermi-Dirac statistics of electrons.
\begin{figure}[ptb]
\begin{center}
\includegraphics[angle=-90, width=0.25\textwidth]
{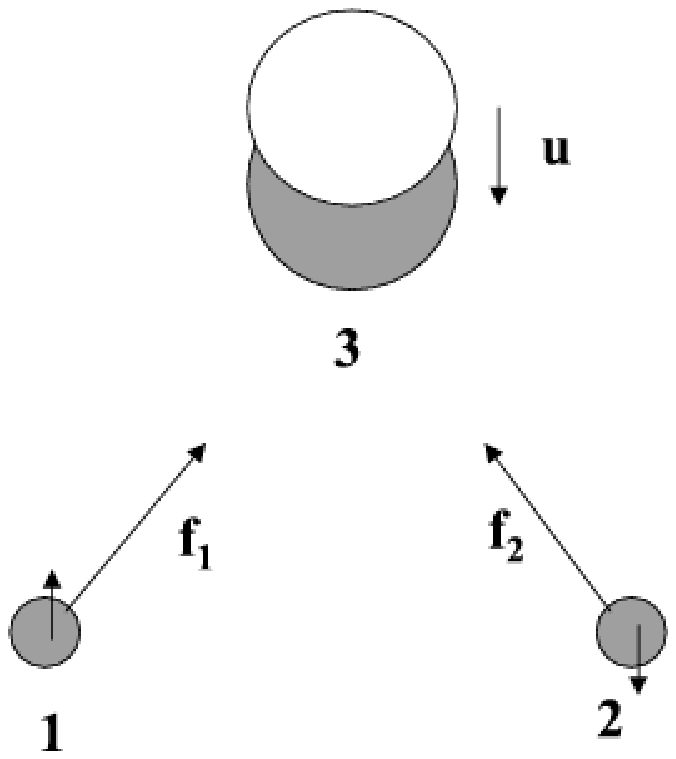}
\vskip-0.5mm
\end{center}
\caption{Two localized electrons at sites $\mathbf{1}$ and
$\mathbf{2}$ shift the equilibrium position of the ion at site
$\mathbf{3}$. As a result, the two
electrons \emph{attract} each other.}%
\label{fig:2sitepol}
\end{figure}

We start with a simple model that illustrates an absence of
switching in molecular quantum dot, which has non-degenerate ($d=1$)
or double-degenerate ($d=2$) level \cite{ABrat06}.
 First, we shall illustrate the failure of the mean-field
approximation on a simplest model of a single atomic level coupled
with a single one-dimensional oscillator with a displacement $x$,
described by a simple Hamiltonian,
\begin{equation}
H=\varepsilon_{0}\hat{n}+fx\hat{n}-{\frac{1}{{2M}}}{\frac{\partial^{2}%
}{{\partial x^{2}}}}+{\frac{kx^{2}}{{2}}}.
\end{equation}
Here $M$ and $k$ are the oscillator mass and the spring constant,
$f$ is the interaction force, and $\hbar=c=k_{B}=1$. This
Hamiltonian is readily diagonalized with the \emph{exact}
displacement transformation of the vibration coordinate $x$,
\begin{equation}
x=y-\hat{n}f/k,
\end{equation}
to the transformed Hamiltonian without electron-phonon coupling,
\begin{align}
\tilde{H}  & =\varepsilon\hat{n}-{\frac{1}{{2M}}}{\frac{\partial^{2}%
}{{\partial y^{2}}}}+{\frac{ky^{2}}{{2}}},\label{eq:newH}\\
\varepsilon & =\varepsilon_{0}-E_{p},\label{eq:newLev}%
\end{align}
where we used $\hat{n}^{2}=\hat{n}$ because of the Fermi-Dirac
statistics. It describes a small polaron at the atomic level
$\varepsilon_{0}$ shifted down by the polaron level shift
$E_{p}=f^{2}/2k$, and entirely decoupled from ion vibrations. The
ion vibrates near a new equilibrium position, shifted by $f/k$, with
the ``old'' frequency $(k/M)^{1/2}$. As a result of the local ion
deformation, the total energy of the whole system decreases by
$E_{p}$ since a decrease of the electron energy by $-2E_{p}$
overruns an increase of the deformation energy $E_{p}$. The major
error of the mean-field approximation of Ref.~\cite{ratsw}
originates in illegitimate replacement of the square of the
occupation number operator $\hat{n}=c_{0}^{\dagger}c_{0}$ by its
``mean-field'' expression $\hat{n}^{2}=n_0\hat{n}$ which contains
the average population of a single molecular level, $n_{0}$, in
disagreement with the exact identity, $\hat{n}^{2}=\hat{n}$. This
leads to a spurious \emph{self-interaction} of a single polaron with
itself [i.e. the term $\varepsilon=\varepsilon_{0}-n_{0}E_{p}$
instead of Eq.~(\ref{eq:newLev})], and a resulting non-existent
nonlinearity in the rate equation.

Lattice deformation also strongly affects the interaction between
electrons. When a short-range deformation potential and molecular
e-ph interactions are taken into account together with the
long-range Fr\"{o}hlich interaction, they can overcome the Coulomb
repulsion. The resulting interaction becomes attractive at a short
distance comparable to a lattice constant. The origin of the
attractive force between two small polarons can be readily
understood from a similar Holstein-like toy model as above
\cite{alebook}, but with two electrons on neighboring sites
\textbf{1,2} interacting with an ion \textbf{3} between them,
Fig.~\ref{fig:2sitepol}. For generality, we now assume that the ion
is a three-dimensional oscillator described by a displacement vector
$\mathbf{u}$, rather than by a single-component displacement $x$ as
in Eq.(1).

The vibration part of the Hamiltonian in the model is
\begin{equation}
H_{ph}=-{\frac{1}{{2M}}\frac{\partial^{2}}{{\partial\mathbf{u}}^{2}}}%
+{\frac{k\mathbf{u}^{2}}{{2}}},\label{eq:Hv}%
\end{equation}
Electron potential energies due to the Coulomb interaction with the
ion are about
\begin{equation}
V_{1,2}=V_{0}(1-\mathbf{u}\cdot\mathbf{e_{1,2}}/a),
\end{equation}
where $\mathbf{e_{1,2}}$ are units vectors connecting sites
$\mathbf{1,2}$ and site $\mathbf{3}$, respectively. Hence, the
Hamiltonian of the model is given by
\begin{equation}
H=E_{a}(\hat{n}_{1}+\hat{n}_{2})+\mathbf{u}\cdot(\mathbf{f_{1}}\hat{n}%
_{1}+\mathbf{f_{2}}\hat{n}_{2})-{\frac{1}{{2M}}\frac{\partial^{2}}%
{{\partial\mathbf{u}}^{2}}}+{\frac{k\mathbf{u}^{2}}{{2}}},
\end{equation}
where $\mathbf{f_{1,2}}=Ze^{2}\mathbf{e_{1,2}}/a^{2}$ is the Coulomb
force, and $\hat{n}_{1,2}$ are occupation number operators at every
site. This Hamiltonian is also readily diagonalized by the same
displacement transformation of the vibronic coordinate $\mathbf{u}$
as above,
\begin{equation}
\mathbf{u}=\mathbf{v}-\left(
\mathbf{f_{1}}\hat{n}_{1}+\mathbf{f_{2}}\hat {n}_{2}\right)  /k.
\end{equation}
The transformed Hamiltonian has no electron-phonon coupling,
\begin{equation}
\tilde{H}=(\varepsilon_{0}-E_{p})(\hat{n}_{1}+\hat{n}_{2})+V_{ph}\hat{n}%
_{1}\hat{n}_{2}-{\frac{1}{{2M}}\frac{\partial^{2}}{{\partial\mathbf{v}}^{2}}%
}+{\frac{k\mathbf{v}^{2}}{{2}}},
\end{equation}
and it describes two small polarons at their atomic levels shifted
by the polaron level shift $E_{p}=f_{1,2}^{2}/2k$, which are
entirely decoupled from ion vibrations. As a result, the lattice
deformation caused by two electrons leads to an effective
interaction between them, $V_{ph}$, which should be added to their
Coulomb repulsion, $V_{c}$,
\begin{equation}
V_{ph}=-\mathbf{f_{1}}\cdot\mathbf{f_{2}}/k.
\end{equation}
When $V_{ph}$ is negative and larger by magnitude than the positive
$V_{c},$ the resulting interaction becomes attractive. That is
$V_{ph}$ rather than $E_{p}$, which is responsible for the
hysteretic behavior of MQDs, as discussed below.

\section{ Exact solution of polaron switching }

The procedure, which fully accounts for all correlations in MQD is
as follows, see Ref.~\cite{AB03vibr}. The molecular Hamiltonian
includes the Coulomb repulsion, $U^{C}$, and the electron-vibron
interaction as
\begin{align}
& H
=\sum_{_{\mu}}\varepsilon_{_{\mu}}\hat{n}_{_{\mu}}+\frac{1}{2}\sum_{_{\mu
}\neq\mu^{\prime}}U_{\mu\mu^{\prime}}^{C}\hat{n}_{_{\mu}}\hat{n}_{\mu^{\prime
}}\nonumber\\
& +\sum_{\mu,q}\hat{n}_{_{\mu}}\omega_{q}(\gamma_{\mu
q}d_{q}+H.c.)+\sum _{q}\omega_{q}(d_{q}^{\dagger}d_{q}+1/2).
\end{align}
Here $d_{q}$ annihilates phonons, $\omega_{q}$ is the phonon
(vibron) frequency, and $\gamma_{\mu q}$ are the electron-vibron
coupling constant ($q$ enumerates the vibron modes). This
Hamiltonian conserves the occupation numbers of molecular states
$\hat{n}_\mu$.

One can apply the canonical unitary transformation $e^{S}$, with
$$
S=-\sum_{q,\mu}\hat{n}_{\mu}\left(  \gamma_{\mu q}d_{q}-H.c.\right)
$$
integrating phonons out. The electron and phonon operators are
transformed as
\begin{equation}
\tilde{c}_{\mu}=c_{\mu}X_{\mu},\qquad X_{\mu}=\exp\left(
\sum_{q}\gamma_{\mu q}d_{q}-H.c.\right)
\end{equation}
and
\begin{equation}
\tilde{d}_{q}=d_{q}-\sum_{\mu}\hat{n}_{\mu}\gamma_{\mu q}^{\ast},
\end{equation}
respectively. This Lang-Firsov transformation shifts ions to new
equilibrium positions with no effect on the phonon frequencies. The
diagonalization is \emph{exact}:
\begin{equation}
\tilde{H}=\sum_{i}\tilde{\varepsilon}_{_{\mu}}\hat{n}_{\mu}+\sum_{q}\omega
_{q}(d_{q}^{\dagger}d_{q}+1/2)+{\frac{1}{{2}}}\sum_{\mu\neq\mu^{\prime}}%
U_{\mu\mu^{\prime}}\hat{n}_{\mu}\hat{n}_{\mu^{\prime}},
\label{eq:HLang}
\end{equation}
where
\begin{equation}
U_{\mu\mu^{\prime}}\equiv
U_{\mu\mu^{\prime}}^{C}-2\sum_{q}\gamma_{\mu
q}^{\ast}\gamma_{\mu^{\prime}q}\omega_{q},\label{eq:Umm1}%
\end{equation}
is the renormalized interaction of polarons comprising their
interaction via molecular deformations (vibrons) and the original
Coulomb repulsion, $U_{\mu\mu^{\prime}}^{C}$. The molecular energy
levels are shifted by the polaron level-shift due to the deformation
created by the polaron,
\begin{equation}
\tilde{\varepsilon}_{_{\mu}}=\varepsilon_{_{\mu}}{-}\sum_{q}|\gamma_{\mu
q}|^{2}\omega_{q}.\label{eq:eps}%
\end{equation}

If we assume that the coupling to the leads is weak, so that the
level width $\Gamma \ll $ $|U|,$ we can find the current from
\cite{meir92}
\begin{eqnarray}
I(V) &=&e\Gamma\int_{-\infty }^{\infty }d\omega \left[ f_{1}(\omega
)-f_{2}(\omega )\right] \rho (\omega ),  \label{eq:Irho} \\
\rho (\omega ) &=&-\frac{1}{\pi }\sum_{\mu }{\rm Im}\hat{G}_{\mu
}^{R}(\omega ),
\end{eqnarray}
where $\left| \mu \right\rangle $ is a complete set of one-particle
molecular states, $f_{1(2)}(\omega)= \left( \exp\frac{\omega+\Delta
\mp eV/2}{T}+1\right)^{-1}$ the Fermi function. Here  $\rho (\omega
)$ is the molecular DOS, $\hat{G}_{\mu }^{R}(\omega )$ is the
Fourier transform of the Green's function $\hat{G}_{\mu }^{R}(t)=$
$-i\theta (t)\left\langle \left\{ c_{\mu }(t),c_{\mu }^{\dagger
}\right\} \right\rangle ,$ $\left\{ \cdots ,\cdots
\right\} $ is the anticommutator, $c_{\mu }(t)=e^{iHt}c_{\mu }e^{-iHt},$ $%
\theta (t)=1$ for $t>0$ and zero otherwise. We calculate $\rho
(\omega )$ {\em exactly} for the Hamiltonian (\ref{eq:HLang}), which
includes both the Coulomb $U^{C}$ and electron-vibron interactions.

The retarded GF becomes
\begin{eqnarray}
G_{\mu }^{R}(t) &=&-i\theta (t)[\left\langle c_{\mu }(t)c_{\mu
}^{\dagger }\right\rangle \left\langle X_{\mu }(t)X_{\mu }^{\dagger
}\right\rangle
\nonumber \\
&&+\left\langle c_{\mu }^{\dagger }c_{\mu }(t)\right\rangle
\left\langle X_{\mu }^{\dagger }X_{\mu }(t)\right\rangle ].
\end{eqnarray}
The phonon correlator is simply
\begin{eqnarray}
\left\langle X_{\mu }(t)X_{\mu }^{\dagger }\right\rangle  &=&\exp \sum_{q}%
\frac{|\gamma _{\mu q}|^{2}}{\sinh \frac{\beta \omega _{q}}{2}}  \nonumber \\
&&\times \left[ \cos \left( \omega t+i\frac{\beta \omega
_{q}}{2}\right) -\cosh \frac{\beta \omega _{q}}{2}\right] ,
\label{eq:Xcorr}
\end{eqnarray}
where the inverse temperature $\beta =1/T$, and $\left\langle X_{\mu
}^{\dagger }X_{\mu }(t)\right\rangle =\left\langle X_{\mu }(t)X_{\mu
}^{\dagger }\right\rangle ^{\ast }.$ The remaining GFs $\left\langle
c_{\mu }(t)c_{\mu }^{\dagger }\right\rangle $, are found from the
equations of motion {\em exactly}. For the simplest case of a
coupling to a single mode with the characteristic frequency $\omega
_{0}$, $\gamma _{q}\equiv \gamma $ and $U_{\mu \mu'}=U$ one obtains
\cite{AB03vibr}
\begin{eqnarray}
G_{\mu }^{R}(\omega ) &=&{\cal
Z}\sum_{r=0}^{d-1}C_{r}(n)\sum_{l=0}^{\infty
}I_{l}\left( \xi \right)   \nonumber \\
&&\biggl[e^{\frac{\beta \omega _{0}l}{2}}\left( \frac{1-n}{\omega
-rU-l\omega _{0}+i\delta }+\frac{n}{\omega -rU+l\omega _{0}+i\delta
}\right)
\nonumber \\
&&+(1-\delta _{l0})e^{-\frac{\beta \omega _{0}l}{2}}  \nonumber \\
&&\times \biggl(\frac{1-n}{\omega -rU+l\omega _{0}+i\delta
}+\frac{n}{\omega -rU-l\omega _{0}+i\delta }\biggr )\biggr],
\label{eq:GwT}
\end{eqnarray}
where
\begin{equation}
{\cal Z}=\exp \left( -\sum_{{\bf q}}|\gamma _{q}|^{2}\coth
\frac{\beta \omega _{q}}{2}\right)
\end{equation}
is the familiar {\em polaron narrowing} factor, the degeneracy
factor
\begin{equation}
C_{r}(n)=\frac{(d-1)!}{r!(d-1-r)!}n^{r}(1-n)^{d-1-r},
\end{equation}
$\xi =|\gamma |^{2}/\sinh \frac{\beta \omega _{0}}{2},$ $I_{l}\left(
\xi \right) $ the modified Bessel function, and $\delta _{lk}$ the
Kroneker symbol.

Then using Eq.(18) the exact spectral function  for a
$d-\mathrm{fold}$ degenerate MQD (i.e. the density of molecular
states, DOS) is found as
\begin{align}
& \rho(\omega)=\mathcal{Z}d\sum_{r=0}^{d-1}C_{r}(n)\sum_{l=0}^{\infty}%
I_{l}\left(  \xi\right) \nonumber\\
& \times\biggl[e^{\beta\omega_{0}l/2}\left[
(1-n)\delta(\omega-rU-l\omega
_{0})+n\delta(\omega-rU+l\omega_{0})\right] \nonumber\\
& +(1-\delta_{l0})e^{-\beta\omega_{0}l/2}[n\delta(\omega-rU-l\omega
_{0})\nonumber\\
& +(1-n)\delta(\omega-rU+l\omega_{0})]\biggr].\label{eq:rho}%
\end{align}

The important feature of DOS, Eq.~(\ref{eq:rho}), is its nonlinear
dependence on the average electronic population $n=\left\langle
c_{\mu}^{\dagger}c_{\mu }\right\rangle ,$ which leads to the
switching, hysteresis, and other nonlinear effects in I-V
characteristics for $d>2$. It appears due to \emph{correlations}
between \emph{different} electronic states via the correlation
coefficients $C_{r}(n)$. There is no nonlinearity if the dot is
nondegenerate, $d=1,$ since $C_{0}(n)=1$. In this simple case the
DOS, Eq.~(\ref{eq:rho}), is a \emph{linear} function of the average
population that can be found as a textbook example of an exactly
solvable problems \cite{mahan}.


In the present case of MQD weakly coupled with leads, one can apply
the Fermi-Dirac golden rule to obtain an equation for $n.$ Equating
incoming and outgoing numbers of electrons in MQD per unit time we
obtain the self-consistent equation for the level occupation $n$ as
\begin{eqnarray}
&&(1-n)\int_{-\infty }^{\infty }d\omega \left\{ \Gamma
_{1}f_{1}(\omega )+\Gamma _{2}f_{2}(\omega )\right\} \rho (\omega )
\nonumber \\
&=&n\int_{-\infty }^{\infty }d\omega \left\{ \Gamma
_{1}[1-f_{1}(\omega )]+\Gamma _{2}[1-f_{2}(\omega )]\right\} \rho
(\omega ), \label{eq:neq}
\end{eqnarray}
where $\Gamma _{1(2)}$ are the transition rates from left (right)
leads to MQD, and $\rho (\omega )$ is found from Eq.(24). For
$d=1,2$ the kinetic equation for $n$ has only one physical root, and
the switching is {\em absent.} Switching appears for $d\geq 3,$ when
the kinetic equation becomes non-linear. Taking into account that
$\int_{-\infty }^{\infty }\rho (\omega )=d$, Eq.~(\ref{eq:neq}) for
the symmetric leads, $\Gamma _{1}=\Gamma _{2},$ reduces to
\begin{equation}
2nd=\int d\omega \rho \left( \omega \right) \left(
f_{1}+f_{2}\right) ,
\end{equation}
which automatically satisfies $0\leq n \leq 1$. Explicitly, the
self-consistent equation for the occupation number is
\begin{equation}
n=\frac{1}{2}\sum_{r=0}^{d-1}Z_{r}(n)[na_{r}+(1-n)b_{r}],
\label{eq:neq1}
\end{equation}
where
\begin{eqnarray}
a_{r}^+ &=&{\cal Z}\sum_{l=0}^{\infty }I_{l}\left( \xi \right) \biggr(e^{\frac{%
\beta \omega _{0}l}{2}}[f_{1}(rU-l\omega _{0})+f_{2}(rU-l\omega
_{0})]
\nonumber \\
&&+(1-\delta _{l0})e^{-\frac{\beta \omega
_{0}l}{2}}[f_{1}(rU+l\omega
_{0})+f_{2}(rU+l\omega _{0})]\biggr),  \label{eq:a} \\
b_{r}^+ &=&{\cal Z}\sum_{l=0}^{\infty }I_{l}\left( \xi \right) \biggr(e^{\frac{%
\beta \omega _{0}l}{2}}[f_{1}(rU+l\omega _{0})+f_{2}(rU+l\omega
_{0})]
\nonumber \\
&&+(1-\delta _{l0})e^{-\frac{\beta \omega
_{0}l}{2}}[f_{1}(rU-l\omega _{0})+f_{2}(rU-l\omega _{0})]\biggr).
\label{eq:b}
\end{eqnarray}
The current is expressed as
\begin{equation}
j\equiv \frac{I(V)}{I_0}=\sum_{r=0}^{d-1}Z_{r}(n)[na_{r}^{-
}+(1-n)b_{r}^{-}],
\end{equation}
where
\begin{eqnarray}
a_{r}^{-} &=&{\cal Z}\sum_{l=0}^{\infty }I_{l}\left( \xi \right) %
\biggr(e^{\frac{\beta \omega _{0}l}{2}}[f_{1}(rU-l\omega
_{0})-f_{2}(rU-l\omega _{0})]  \nonumber \\
&&+(1-\delta _{l0})e^{-\frac{\beta \omega
_{0}l}{2}}[f_{1}(rU+l\omega
_{0})-f_{2}(rU+l\omega _{0})]\biggr),  \label{eq:at} \\
b_{r}^{- } &=&{\cal Z}\sum_{l=0}^{\infty }I_{l}\left( \xi \right) %
\biggr(e^{\frac{\beta \omega _{0}l}{2}}[f_{1}(rU+l\omega
_{0})-f_{2}(rU+l\omega _{0})]  \nonumber \\
&&(1-\delta _{l0})e^{-\frac{\beta \omega _{0}l}{2}}[f_{1}(rU-l\omega
_{0})-f_{2}(rU-l\omega _{0})]\biggr), \label{eq:bt}
\end{eqnarray}
and $I_0=ed\Gamma$.

\section{ Absence of switching of single- or double-degenerate MQD
}

\begin{figure}[ptb]
\begin{center}
\includegraphics[angle=0, width=0.4\textwidth]
{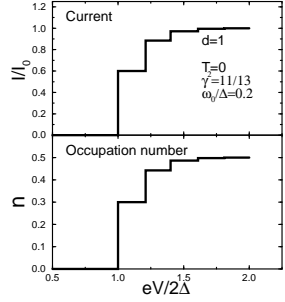}
\end{center}
\caption{Current-voltage characteristic of the nondegenerate ($d=1)$
MQD at $T=0,$ $\omega_{0}/\Delta=0.2$, and $\gamma^{2}=11/13$. There
is the phonon ladder in I-V , but no hysteresis.}
 \label{fig:deg1}
\end{figure}
\begin{figure}[ptb]
\begin{center}
\includegraphics[angle=-0, width=0.4\textwidth]
{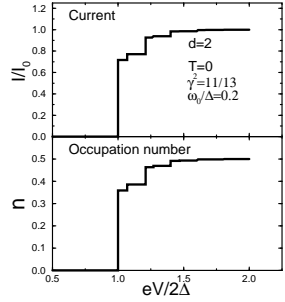}
\end{center}
\caption{Current-voltage characteristic of two-fold degenerate MQDs
($d=2)$ does not show hysteretic behavior. The parameters are the
same as in Fig.~\ref{fig:deg1}. Larger number of elementary
processes for conductance compared to the nondegenerate case of
$d=1$ generates more steps in the phonon ladder in comparison with
Fig.~\ref{fig:deg1}.
\label{fig:deg2} }%
\end{figure}
%

If the transition rates from electrodes to MQD are small,
$\Gamma\ll\omega _{0}$, the rate equation for $n$ and the current,
$I(V)$ are readily obtained by using the exact molecular DOS,
Eq.~(\ref{eq:rho}) and the Fermi-Dirac Golden rule. In particular,
for the nondegenerate MQD and $T=0$K the result is
\begin{equation}
n=\frac{b_{0}^{+}}{2+b_{0}^{+}-a_{0}^{+}},
\end{equation}
and
\begin{equation}
j=\frac{2b_{0}^{-}+a_{0}^{-}b_{0}^{+}-a_{0}^{+}b_{0}^{-}}{2+b_{0}^{+}%
-a_{0}^{+}}.
\end{equation}
The general expressions for the coefficients Eqs.~(28,29) and
Eqs.(31,32) at arbitrary temperatures in Ref.\cite{AB03vibr} are
simplified in low-temperature limit as
\begin{align}
& a_{0}^{\pm}=\mathcal{Z}\sum_{l=0}^{\infty}\frac{|\gamma|^{2l}}{l!}%
[\Theta(l\omega_{0}-\Delta+eV/2)\nonumber\\
& \pm\Theta(l\omega_{0}-\Delta-eV/2)],\\
& b_{0}^{\pm}=\mathcal{Z}\sum_{l=0}^{\infty}\frac{|\gamma|^{2l}}{l!}%
[\Theta(-l\omega_{0}-\Delta+eV/2)\nonumber\\
& \pm\Theta(-l\omega_{0}-\Delta-eV/2)],
\end{align}
where  $\Delta$ is the position of the MQD level with respect to the
Fermi level at $V=0$, and $\Theta(x)=1$ if $x>0$ and zero otherwise.
The current is single valued, Fig.~\ref{fig:deg1}, with the familiar
steps due to phonon-side bands.

On the contrary, the mean-field approximation (MFA) leads to the
opposite conclusion. Galperin \textit{et al}.~\cite{ratsw} have
replaced the occupation number operator $\hat{n}$ in the e-ph
interaction by the average population $n_{0}$ [Eq.~(2) of
Ref.~\cite{ratsw}] and found the average steady-state vibronic
displacement $\langle d+d^{\dagger}\rangle$ proportional to $n_{0} $
(this is an explicit \emph{neglect of all quantum fluctuations }on
the dot accounted for in the exact solution). Then, replacing the
displacement operator $d+d^{\dagger}$ in the bare Hamiltonian,
Eq.~(11), by its average,
Ref.~\cite{ratsw}, they obtained a new molecular level, $\tilde{\varepsilon}%
_{0}=\varepsilon_{0}-2\varepsilon_{reorg}n_{0}$ shifted linearly
with the average population of the level. This is in stark
disagreement with the
conventional constant polaronic level shift, Eq.~(\ref{eq:newLev}%
,\ref{eq:eps}) ($\varepsilon_{reorg}$ is $|\gamma|^{2}\omega_{0}$ in
our notations). The MFA spectral function turned out to be highly
nonlinear as a function of the population, e.g. for the
weak-coupling with the leads
$\rho(\omega)=\delta(\omega-\varepsilon_{0}-2\varepsilon_{reorg}n_{0}),$
see Eq.~(17) in Ref.~\cite{ratsw}. As a result, the authors of
Ref.\cite{ratsw} have found multiple solutions for the steady-state
population, Eq.~(15) and Fig.~1, and switching, Fig.~4 of
Ref.~\cite{ratsw}, which actually do not exist being an artefact of
the approximation.

In the case of a double-degenerate MQD, $d=2,$ there are two terms,
which contribute to the sum over $r$, with $C_{0}(n)=1-n$ and
$C_{1}(n)=n.$ The rate equation becomes a quadratic one
\cite{AB03vibr}. Nevertheless there is only one physical root for
any temperature and voltage, and the current is also single-valued.
 The double-degenerate level provides more elementary
processes for conductance reflected in larger number of steps on
phonon ladder compared to $d=2$ case, Fig.~\ref{fig:deg2}.

Note that the mean-field solution by Galperin \textit{et al.}
\cite{ratsw} applies at any ratio $\Gamma/\omega_{0},$\ including
the limit of interest to us, $\Gamma\ll\omega_{0}.$ where their
transition between the states with $n_{0}=0$ and $1$ only sharpens,
but none of the results change. Therefore, MFA predicts a current
bistability in the system where it does not exist at $d=1.$ Ref.
\cite{ratsw} plots the results for $\Gamma\geq\omega_{0},$
$\Gamma\approx0.1-0.3$ eV, which corresponds to molecular bridges
with a resistance of about a few $100$K$\Omega.$ Such model
``molecules'' are rather ``metallic'' in their conductance and could
hardly show any bistability at all because carriers do not have time
to interact with vibrons on the molecule. Indeed, taking into
account the coupling with the leads beyond the second order and the
coupling between the molecular and bath phonons could hardly provide
any non-linearity because these couplings do not depend on the
electron population. This rather obvious conclusion for molecules
strongly coupled to the electrodes can be reached in many ways, see
e.g. a derivation in Refs.~\cite{millis05,mozyr06}. While Refs.
\cite{millis05,mozyr06} do talk about telegraph current noise in the
model, there is no hysteresis in the adiabatic regime,
$\Gamma\gg\omega_{0}$ either. This result certainly has nothing to
do with our mechanism of switching \cite{AB03vibr} that applies to
molecular quantum dots ($\Gamma\ll\omega_{0})$ with $d>2.$ Such
regime has not been studied in
Refs.~\cite{millis05,mozyr06,millis04}, which have applied the
adiabatic approximation, as being ``too challenging problem''.
Nevertheless, Mitra \textit{et al}. \cite{millis04} have
misrepresented our formalism \cite{AB03vibr} claiming that it
``lacks of renormanlization of the dot-lead coupling'' (due to
electron-vibron interaction), or ``treats it in an average manner''.
In fact, the formalism \cite{AB03vibr} is exact, fully taking into
account the polaronic renormalization, phonon-side bands and
polaron-polaron correlations in the exact molecular DOS,
Eq.~(\ref{eq:rho}).

As a matter of fact, most of the molecules are very resistive, so
the actual molecular quantum dots are in the regime we study, see
Ref.\cite{mqdexp}. For example, the resistance of fully conjugated
three-phenyl ring Tour-Reed molecules chemically bonded to metallic
Au electrodes \cite{ReedRAM} exceeds $1$G$\Omega$. Therefore, most
of the molecules of interest to us are in the regime that we
discussed, not that of Refs.\cite{millis05,mozyr06}.

\section{Nonlinear rate equation and switching}

The  switching appears only for $d>2$. For example, for $d=4$
 the rate equation (27) is of the fourth power in $n,$%
\begin{eqnarray}
2n &=&(1-n)^{3}[na_{0}^+ +(1-n)b_{0}^+]  \nonumber \\
&&+3n(1-n)^{2}[na_{1}^+ +(1-n)b_{1}^+]  \nonumber \\
&&+3n^{2}(1-n)[na_{2}^+ +(1-n)b_{2}^+]  \nonumber \\
&&+n^{3}[na_{3}^+ +(1-n)b_{3}^+].  \label{eq:nab}
\end{eqnarray}

\begin{figure}[tbp]
\begin{center}
\includegraphics[angle=-0,width=0.67\textwidth]{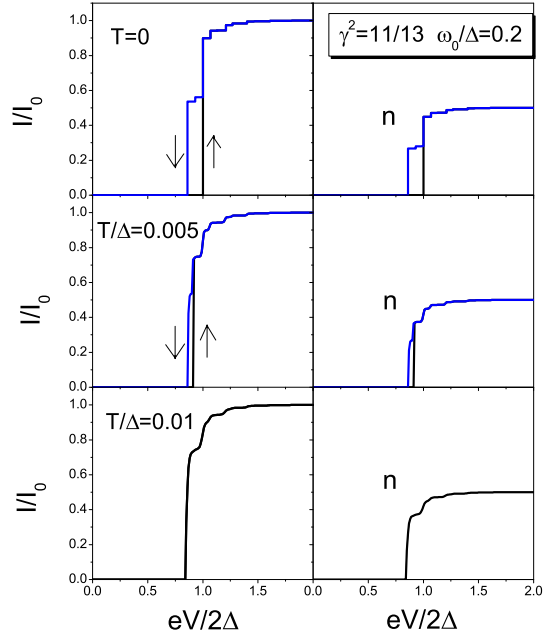} \vskip -0.5mm
\end{center}
\caption{The bistable I-V curves for tunneling through molecular
quantum dot  with the electron-vibron coupling constant
$\gamma^2=11/13$ and $\omega_0/\Delta = 0.2$. The up arrows show
that the current picks up at some voltage when it is biased, and
then drops at lower voltage when the bias is being reduced. The bias
dependence of current basically repeats the shape of the level
occupation $n$ (right column). Steps on the curve correspond to the
changing population of the phonon side-bands}
\end{figure}
Different from the non-degenerate or double-degenerate MQD, the rate
equation  for $d=4$
  has two stable
 physical roots in a certain voltage range and the current-voltage
characteristics show a hysteretic behavior. Our
 numerical results \cite{AB03vibr} for $\omega _{0}=0.2$ (in units of $\Delta
,$ as all the energies in the problem), $U^{C}=0$, and for the
coupling constant, $\gamma ^{2}=11/13$ are shown in Fig.5. This case
formally corresponds to a negative Hubbard $U=-2\gamma ^{2}\omega
_{0}\approx -0.4$  (we selected those values of $\gamma^2$ to avoid
accidental commensurability of  the correlated levels separated by
$U$ and the phonon side-bands). The threshold for the onset of
bistability appears at a voltage bias $eV/2\Delta =0.86 $ for
$\gamma ^{2}=11/13$ and $\omega _{0}=0.2$. The steps on the I-V
curve, Fig.5, are generated by the phonon side-bands originating
from correlated levels in the
dot with the energies $\Delta ,$ $\Delta +U,$ ..., $\Delta +(d-1)U.$ Since $%
\omega _{0}$ is not generally commensurate with $U,$ we obtain quite
irregular picture of the steps in I-V\ curves. The bistability
region shrinks down with temperature.

Note that switching required a degenerate MQD ($d>2)$ and the weak
coupling to the electrodes, $\Gamma \ll \omega _{0}.$ Different from
the non-degenerate dot, the rate equation for a multi-degenerate
dot, $d>2$, weakly coupled to the leads has multiple physical roots
in a certain voltage range and a hysteretic behavior due to {\em
correlations} between different electronic states of MQD.

\section{Summary}

We have calculated the I-V characteristics of a nondegenerate
($d=1$), two-fold degenerate ($d=2$) molecular quantum dots showing
no hysteretic behavior of current, and concluded that mean field
approximation \cite{ratsw} leads to a non-existent swtching in a
model that was solved exactly in Ref.~\cite{AB03vibr}. Different
from the non-degenerate and two-fold degenerate dots, the rate
equation for a multi-degenerate dot, $d>2$, weakly coupled to the
leads, has multiple physical roots in a certain voltage range
showing hysteretic behavior due to \emph{correlations} between
different electronic states of MQD \cite{AB03vibr}. Pair tunneling
is also allowed in out model, though it should only result in tiny
peaks on the background of the main current contributed by single
polaron tunneling. Our conclusions are important for searching for
the current-controlled polaronic molecular-size switches.
Incidentally, C$_{60}$ molecules have the degeneracy $d=6$ of the
lowest unoccupied level, which makes them one of the most promising
candidate systems, if the weak-coupling with leads is secured.

We thank the participants of the ESF workshop "Mott's Physics in
Nanowires and Quantum Dots" (Cambridge, UK, 31 July-2 August, 2006)
for many discussions, and greatly appreciate the  financial support
of the European Science Foundation (ESF) and EPSRC (UK) (grants no
EP/C518365/1 and no EP/D07777X/1).



%
%
%
%
%
%



\printindex
\end{document}